\documentclass[%
 reprint,
 amsmath,amssymb,
 aps,
]{revtex4-2}

\usepackage{graphicx}
\usepackage{dcolumn}
\usepackage{bm}
\usepackage{color}
\usepackage[normalem]{ulem}
\usepackage{braket}

\usepackage{comment}
\usepackage{mathtools}

\begin{document}

\preprint{APS/123-QED}

\title{Quantum-Accelerated Solution of Nonlinear Equations from Variational Principles}

\author{Katsuhiro Endo}
 \email{katsuhiro.endo@aist.go.jp}
\author{Kazuaki Z. Takahashi}%
 \email{kazu.takahashi@aist.go.jp}
\affiliation{%
 National Institute of Advanced Industrial Science and Technology (AIST),\\
 Materials DX Research Center,\\
 Central 2, 1-1-1 Umezono, Tsukuba, Ibaraki 305-8568, Japan
}%

\date{\today}

\begin{abstract}
Nonlinear equilibrium problems derived from variational principles arise throughout physics and engineering, including structural mechanics, fluid dynamics, and electromagnetism. While fault-tolerant quantum algorithms have shown promising advantages for linear systems and linear dynamical simulations, extending quantum acceleration to nonlinear equilibrium problems remains a major challenge.
Here we introduce a quantum algorithmic framework for nonlinear equilibrium analysis based on gradient-flow linearization. The key idea is to reformulate equilibrium conditions as nonlinear gradient-flow dynamics and transform the resulting evolution into a linear dynamical system using exact linearization techniques such as Carleman and Pivot Switching Carleman (PSC) linearization. This construction enables nonlinear equilibrium and energy-minimization problems to be addressed using quantum algorithms for linear dynamical simulation.
We demonstrate the framework for nonlinear elasticity problems ranging from single nonlinear springs and chain-spring systems to two-dimensional truss structures. The resulting truncated linearized dynamics accurately reproduce nonlinear equilibrium states, while PSC linearization substantially improves stability in regimes where conventional Carleman linearization becomes unreliable.
More broadly, our work establishes a connection between variational principles, nonlinear energy minimization, exact linearization, and quantum dynamical simulation. This perspective opens a route toward quantum algorithms for nonlinear physical systems beyond the scope of existing linear-system-based approaches.
\end{abstract}

\maketitle

\begin{figure*}[t]
\includegraphics[width=1.0\textwidth]{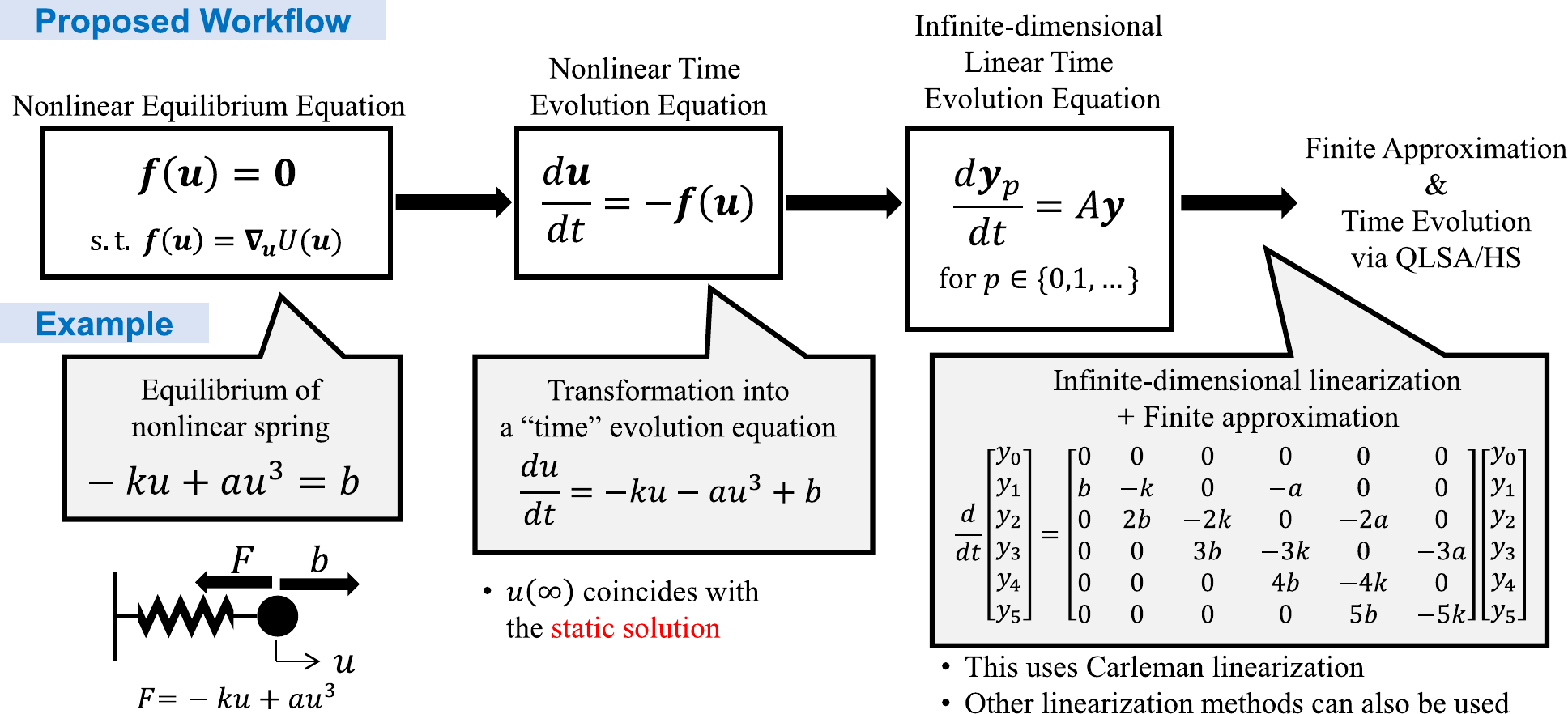}
\caption{Flowchart of the proposed method for transforming a nonlinear equilibrium equation into a linear system solvable by quantum algorithms. The nonlinear equilibrium equation is first reformulated as a nonlinear time-evolution equation. The time variable in this gradient-flow dynamics is an auxiliary parameter and does not necessarily represent the physical time of the original equilibrium problem. The time-evolution equation is then exactly linearized (e.g., using Carleman linearization or the PSC method) into an infinite-dimensional system. Subsequently, a finite-dimensional approximation is applied, enabling the problem to be solvable using a QLSA or HS.}
\label{fig:schema}
\end{figure*}

\section{\label{sec:intro}Introduction}

The variational principle, which is most prominently expressed as the principle of least action, is a foundational and unifying concept in theoretical physics. It provides a systematic and elegant framework for describing the dynamics of a broad class of physical systems, from classical mechanics \cite{landau2013course} to quantum field theory \cite{feynman2005principle} and general relativity \cite{Hilbert1915}. Formulating physical laws as the extremization of an action integral yields the governing equations of motion and reveals deep connections between symmetries and conservation laws via Noether’s theorem \cite{noether1983invariante}. This formulation is remarkably general: it applies to both discrete and continuous systems and bridges classical and quantum descriptions—most notably through path-integral formalism. More broadly, variational principles can be constructed systematically for very general prescribed equations, as formalized in Refs.~\cite{RevModPhys.55.725,rau1976cross,PhysRevA.18.971}.
 
Beyond fundamental physics, the variational principle underpins many classical theories through the extremization of energy- or entropy-based functionals. In solid mechanics, the principle of minimum potential energy governs the equilibrium of elastic structures and forms the basis of numerical methods such as the finite element method \cite{washizu1968variational}. In fluid dynamics, the Euler equations for ideal flows are derived using Hamilton’s principle \cite{morrison1998hamiltonian}. The electromagnetic theory expresses Maxwell’s equations through a variational formalism involving the electromagnetic field tensor and Lagrangian density \cite{jackson2021classical}. In thermodynamics and statistical mechanics, maximizing entropy yields equilibrium distributions such as the Gibbs ensemble \cite{landau1981statistical}. These diverse applications demonstrate the versatility of the principle in capturing nonlinear phenomena, including large deformations, turbulent flows, nonlinear wave propagation, and temperature-dependent transport.
 
Solving the nonlinear equations resulting from these variational formulations typically relies on classical computing. Sophisticated numerical techniques, including nonlinear finite element analysis, Newton--Raphson solvers, and multiscale methods, enable high-fidelity simulations of complex physical systems \cite{tadmor2012review}. However, in strongly nonlinear regimes, classical approaches encounter persistent challenges such as high computational costs, convergence failure, sensitivity to parameters, and the presence of bifurcations or stochastic effects \cite{knoll2004jacobian,de2021uncertainty}. These limitations have motivated the exploration of emerging computational strategies such as reduced-order modeling, uncertainty quantification, and quantum computing.
 
Fault-tolerant quantum computers (FTQCs) are anticipated to offer computational advantages for solving problems that are intractable in classical machines. Following early breakthroughs such as Shor’s and Grover’s algorithms \cite{shor1994algorithms,grover1996fast}, recent efforts have explored the quantum acceleration of classical simulations, particularly for linear differential equations such as heat and wave equations \cite{linden2022quantum,costa2019quantum}. These results suggest that FTQCs could become powerful tools in simulation-driven science and engineering.
 
Quantum acceleration algorithms can be broadly categorized into two axes: equilibrium versus dynamical analysis, and linear versus nonlinear systems. For linear equilibrium problems, various quantum linear system algorithms (QLSAs) have been developed, including the Harrow—Hassidim--Lloyd (HHL) algorithm \cite{harrow2009quantum}, quantum singular value transformation (QSVT) \cite{gilyen2019quantum}, and adiabatic methods \cite{costa2022optimal,subacsi2019quantum}. These methods have been applied to systems resulting from finite element discretizations \cite{montanaro2016quantum} and electromagnetic simulations\cite{sinha2010quantum}. To address scaling issues related to large condition numbers, quantum preconditioning techniques have been introduced, as demonstrated by Clader et al. \cite{clader2013preconditioned}. For linear dynamical systems, both QLSA-based \cite{berry2017quantum,berry2014high} and Hamiltonian simulation (HS)-based \cite{engel2019quantum,novikau2022quantum,miyamoto2024quantum,costa2019quantum,suau2021practical} approaches have been proposed. Babbush et al. demonstrated an exponential quantum advantage in simulating large-scale coupled oscillator systems using HSs \cite{babbush2023exponential}. Moreover, several studies have extended HSs to non-unitary dynamics using techniques such as unitary dilation \cite{gonzalez2023efficient}, warped phase transformations \cite{jin2022quantum}, and linear combination of unitaries \cite{an2023linear,an2023quantum}. Rau and Wendell earlier embedded dissipative density-matrix dynamics into a higher-dimensional Liouville--Bloch equation for unitary-integration schemes \cite{PhysRevLett.89.220405}.
 
In contrast, quantum algorithms for nonlinear problems are relatively underdeveloped. For nonlinear dynamical systems, Carleman and Koopman-von Neumann linearization techniques \cite{liu2021efficient,engel2021linear,liu2023efficient,costa2023further,joseph2020koopman,JIN2023112149,krovi2023improved} have been proposed to embed nonlinear evolution into high-dimensional linear systems, which can then be addressed using QLSAs. Liu et al. demonstrated that second-order nonlinear systems can be simulated in this manner \cite{liu2021efficient}. Our prior work introduced the pivot-switching Carleman method \cite{endo2024divergence}, which improves the numerical stability and extends the applicability to strongly nonlinear systems, including phase-field models.

Only a few approaches exist for solving nonlinear equilibrium problems. Xue et al. proposed a homotopy perturbation method to reduce quadratic nonlinearities to a linear form, thereby achieving an exponential speedup in the state dimension \cite{xue2022quantum}. Nghiem et al. developed a quantum version of Newton’s method based on block-encoded arithmetic oracles that can solve general homogeneous nonlinear equations, although it requires exponentially many oracle queries with respect to the steps \cite{nghiem2024quantum}. Despite these promising directions, the practical applications to real-world nonlinear systems remain scarce and largely unvalidated.
 
In this paper, we propose a quantum algorithmic framework for solving nonlinear equilibrium equations derived from variational principles. The key concept is to reformulate the equilibrium condition as nonlinear gradient-flow dynamics and transform the resulting time-evolution equation into a linear dynamical system using exact infinite-dimensional linearization techniques. This construction enables nonlinear equilibrium problems to be addressed using quantum algorithms for linear dynamical simulations on FTQCs.

We demonstrate the applicability of this approach using several examples of nonlinear elasticity, including single nonlinear springs, chain--spring systems, and two-dimensional truss structures. The results show that truncated linearized dynamics can accurately reproduce nonlinear equilibrium states and capture the effects of structural nonlinearity.

We further analyze the computational resources required for the proposed framework and show that, for representative systems, the query complexity of the resulting linearized dynamics can be bounded independently of the system size under a constant evolution time. In addition, we introduce a pre-initialization strategy based on the solution of the corresponding linear equilibrium problem and analyze how this initialization can reduce the convergence time of the gradient-flow dynamics under weak nonlinearity assumptions.

From a broader perspective, the proposed framework can also be interpreted as a quantum approach to nonlinear energy minimization problems. By connecting variational formulations, gradient-flow dynamics, and exact linearization techniques, this paper provides a new pathway toward quantum algorithms for nonlinear physical systems and optimization problems that occur in physics and engineering.

\begin{figure*}[t]
\includegraphics[width=0.9\textwidth]{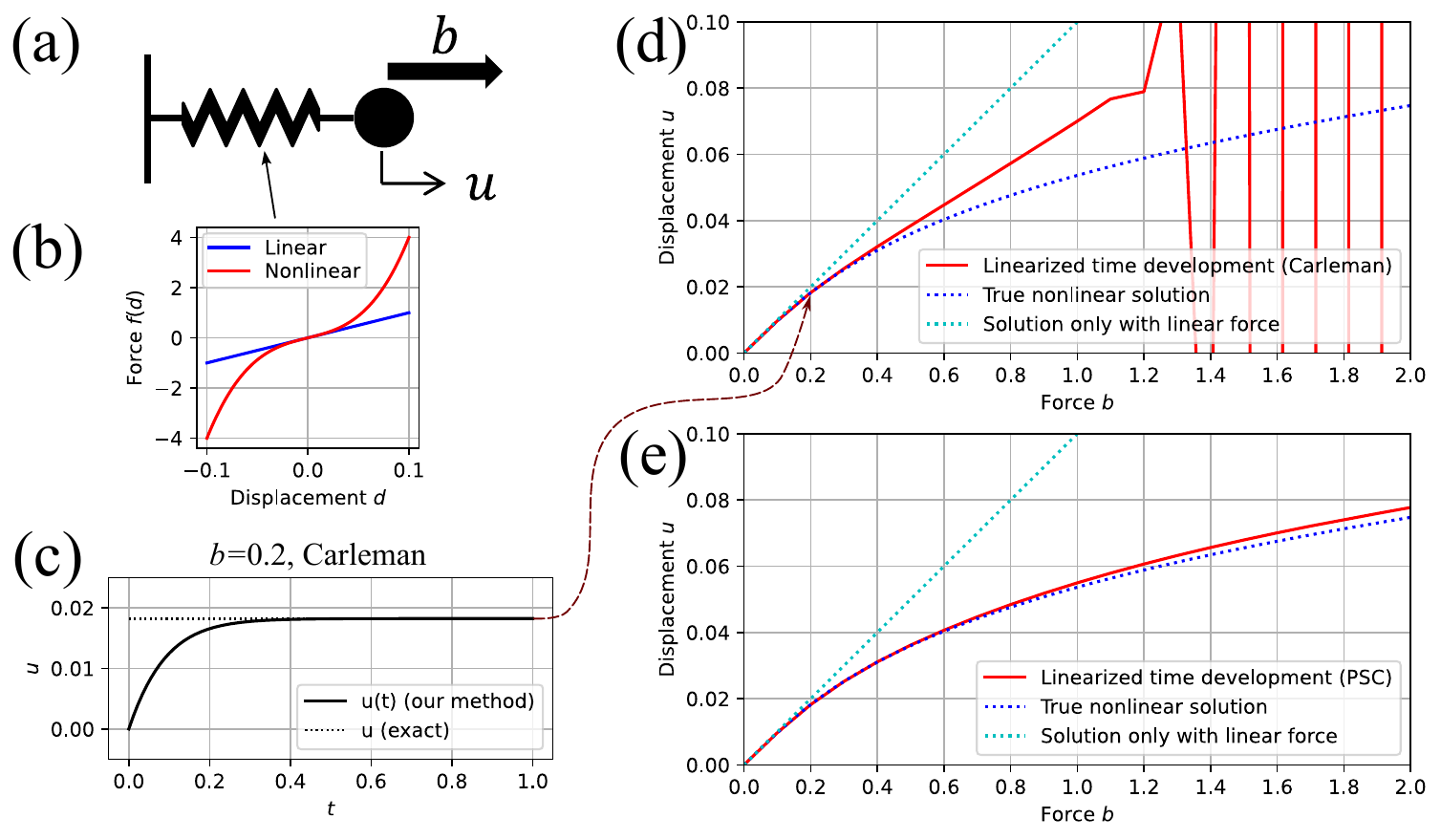}
\caption{Results for a one-dimensional single nonlinear spring system.
(a) Target spring--mass system. (b) Force--displacement relations for the nonlinear spring $k=10,a=3000$ (red) and $k=10, a=0$ (blue).
(c) Example of the time evolution of the displacement $u$ obtained using our method, converging to the correct solution. (d) Displacement obtained using our method with Carleman linearization for each strength of the external force (red). The exact solution (blue dashed) and linear-spring case (light-blue dashed) are included for comparison. (e) Displacement obtained using our method with PSC linearization. }
\label{fig:experimentX}
\end{figure*}

\section{Method}
In this section, we describe the setting of our target problem and our simple method for solving nonlinear equilibrium equations. 

\subsection{Problem Statement}
Let $\bm{u} = (u_1,u_2,\cdots,u_D) \in \mathbb{R}^D$
be a state vector of the system and $\bm{f}(\cdot) : \mathbb{R}^D \rightarrow \mathbb{R}^D$
be a system-specific vector-valued nonlinear function.
We consider the problem of determining an equilibrium point $\bm{u}^*$ satisfying the following equilibrium equation:
\begin{align}
    \bm{f}(\bm{u}^*) = \bm{0}.
    \label{eq:nonlineq}
\end{align}
In the variational principle, the equilibrium equations always have a variational form; that is, a scalar-valued potential function $U(\cdot) : \mathbb{R}^D \rightarrow \mathbb{R}$ exists that satisfies
\begin{align}
     \bm{f}(\bm{u}) = \nabla_{\bm{u}} U(\bm{u}).
     \label{eq:fconserve}
\end{align}

To avoid instability, we assume $U$ satisfies the following two conditions: (1) $\nabla_{\bm{u}}U$ is locally Lipschitz continuous anywhere and (2) $U(\bm{u})\rightarrow \infty$ when $\|\bm{u}\| \rightarrow \infty$.
Our proposed method aims to determine any $\bm{u}^*$ that satisfies Eq.~(\ref{eq:nonlineq}), given that a nonlinear function $\bm{f}$ satisfies the conditions above.

\subsection{Proposed Method}
To solve the nonlinear equilibrium equation Eq.~(\ref{eq:nonlineq}), we consider the following time-evolution equation: 
\begin{align}
    \frac{d\bm{u}}{dt} = - \bm{f}(\bm{u}).
    \label{eq:fdynamicsnonlin}
\end{align}
Here, $t$ denotes an auxiliary time parameter introduced for the gradient-flow dynamics and does not, in general, represent the physical time of the original equilibrium problem.
Starting from any initial state $\bm{u}(0)$ and solving the above time-evolution equation, $\bm{u}(t)$ always reaches a state satisfying Eq.~(\ref{eq:nonlineq}) as $t\rightarrow \infty$ (see Appendix A for a simple proof).
This implies that rather than solving the nonlinear equilibrium equation in Eq.~(\ref{eq:nonlineq}) directly, we solve the equations to perform the time evolution that follows Eq.~(\ref{eq:fdynamicsnonlin}).

Because Eq.~(\ref{eq:fdynamicsnonlin}) is generally nonlinear, we can use the exact infinite-dimensional linearization technique to transform nonlinear time-evolution equations into infinite-dimensional linear time-evolution equations, such as Carleman linearization and Koopman von Neumann linearization. Note that these linearization methods cannot be directly applied to nonlinear equations without time dependence, such as Eq.~(\ref{eq:nonlineq}). Therefore, converting it into a time-dependent form, as shown in Eq.~(\ref{eq:fdynamicsnonlin}) is a simple but important feature of our proposed method. Subsequently, by truncating the resulting infinite-dimensional linear equations to a finite dimension, efficient linear dynamical simulations can be performed using methods such as QLSAs and HSs, as described in the introduction. Fig.~1 shows a flowchart of the proposed method.

\section{\label{sec:results}Examples and Numerical Results}

In this section, we demonstrate the applicability of our method by solving three problems in the field of nonlinear elasticity. First, we examine the simplest nonlinear system: a single-point nonlinear spring model. We formulate truncated linearized time-evolution equations derived from our proposed method and verify the numerical results of a dynamic simulation. Second, we evaluate the computational cost of our method for a one-dimensional multi-mass nonlinear spring model. We also present numerical simulation results under specific conditions for the second model. Third, we demonstrate the extensibility of our method by solving small two-dimensional truss problems.

\subsection{One-dimensional single nonlinear spring}

We consider a one-dimensional space in which a fixed point A located at $x=-1$ and a point mass B located at $x = u$ are connected by a single spring, as shown in Fig.~2(a). The spring has a natural length of 1, and its force for restoring $f(d)$ with respect to the spring displacement $d$ is given by 
\begin{align}
    f(d) = kd + ad^3,
    \label{eq:springforce}
\end{align}
where $k>0$ and $a \geq 0$ characterize the spring properties. Fig.~2(b) shows this force--displacement relationship, which indicates that the spring is made of a nonlinear hyper-elastic material.

When an external force $b \geq 0$ is applied to point mass B, the equilibrium equations for the external force and internal spring forces are obtained as follows:
\begin{align}
    ku + au^3 -b= 0,
    \label{eq:problemeq_exp1}
\end{align}
which is the equation used for the solution. This equation contains a cubic term for the unknown variable $u$ and is classified as a nonlinear equilibrium equation.

To use the proposed method, we can determine the scalar potential 
\begin{align}
    U(u) \coloneqq \frac{1}{2}ku^2 + \frac{1}{4} au^4 - bu,
\end{align}
which satisfies the relationship
\begin{align}
    ku + au^3 - b = \nabla_u U(u).
\end{align}
This relationship can be derived from the fact that $U$ is the total potential energy or the sum of the internal potential energy of the spring and the work done by the external force. According to the principle of stationary potential energy, which is a variational principle, the system must exist at a stationary point ($\nabla_u U=0$) in equilibrium.

The potential $U$ satisfies the conditions required for our proposed method: Any multivariate polynomial function is locally Lipschitz continuous and $U(d)$ approaches infinity as $|u|\rightarrow \infty$.

Therefore, by using the proposed method, the nonlinear equilibrium equation Eq.~(\ref{eq:problemeq_exp1}) can be transformed into the following nonlinear time-evolution equation:
\begin{align}
    \frac{du}{dt} = -ku - au^3 +b.
\end{align}
This time evolution can be implemented on a quantum computer by utilizing exact linearization of infinite-dimensional linear dynamics. In Carleman linearization, an infinite number of variables, denoted by $y_p = u^p$ ($p\in \mathbb{N} \cup \{0\}$), is introduced. The time derivative of $y_p$ results in the dynamics of the infinite-dimensional system:
\begin{align}
    \frac{dy_p}{dt} = p(-ky_p-ay_{p+2}+by_{p-1}).
\end{align}

The resulting equation is a linear time-evolution equation. Truncating all the terms $p> P \in \mathbb{Z}$ yields a truncated finite-dimensional linear time-evolution equation. For example, when $P=5$, this can be expressed in matrix form as
\begin{align}
    \frac{d}{dt}
    \begin{bmatrix}
        y_0 \\ y_1 \\ y_2 \\ y_3 \\ y_4 \\ y_5
    \end{bmatrix} = 
    \begin{bmatrix}
        0 & 0 & 0 & 0 & 0 & 0 \\
        b &-k & 0 &-a & 0 & 0 \\
        0 & 2b&-2k& 0 &-2a& 0 \\
        0 & 0 & 3b&-3k& 0 &-3a\\
        0 & 0 & 0 & 4b&-4k& 0 \\
        0 & 0 & 0 & 0 & 5b&-5k
    \end{bmatrix}
    \begin{bmatrix}
        y_0 \\ y_1 \\ y_2 \\ y_3 \\ y_4 \\ y_5
    \end{bmatrix}.
    \label{eq:exp1CL}
\end{align}
This time-evolution equation can be simulated using well-known quantum algorithms such as QSLAs and HSs.

Next, we verify the solution to the nonlinear equilibrium equation obtained by solving the linear time evolution above. We numerically integrate Eq.~(\ref{eq:exp1CL}) and obtain the value of $y_1(t=T_\mathrm{end})$ for each external force $b=0.0,0.1,\cdots, 2.0$. 
We use $T_\mathrm{end} = 1$ as a sufficiently long time. We also use the spring properties $k=10$ and $a=3000$ in the numerical simulation results. The initial value of the state $u(t=0)=0$ is set as the natural state with no external force. Fig.~2(c) shows an example of the time-evolution trajectory of $y_1(t)$ at $b=0.2$.
In Fig.~2(d), the red line represents $y_1(t=T_\mathrm{end})$ as a solution to Eq.~(\ref{eq:problemeq_exp1}) based on the proposed method and Carleman linearization. The blue dotted line indicates the exact solution of Eq.~(\ref{eq:problemeq_exp1}). Additionally, the light-blue line shows the exact solution for the linear spring case, where we set $k=10, a=0$ in Eq.~(\ref{eq:problemeq_exp1}).

Fig.~2(d) indicates that the use of our method with Carleman linearization successfully replicates the exact solution up to a limit of $b < 0.3$. Even in this range of $b$, both the region where the nonlinear spring behaves as a linear spring ($b \leq 0.1$) and the region where the nonlinearity of the spring appears and the spring displacement is suppressed compared with that of the linear spring ($0.1 < b < 0.3$) can be correctly solved. This implies that the proposed method can solve nonlinear equilibrium equations beyond the range of linearity.

Conversely, for $b \geq 0.3$, the error increases in proportion to the rise in $b$. For $b \geq 1.2$, the time evolution diverged.
Eigenvalue analysis of the system matrix in Eq.~(\ref{eq:exp1CL}) reveals that the real parts become positive when $b>1.145$. Linear dynamics with positive eigenvalues always diverge. Therefore, the solution is essentially invalid for $b$ within this range. We also confirm that this behavior persists even if the truncation degree $P$ increases substantially.

Such divergence problems have been reported in previous studies that employed Carleman linearization \cite{sanavio2024three,itani2022analysis,itani2024quantum}. To improve this problem, our earlier work proposed pivot switching Carleman (PSC) linearization \cite{endo2024divergence}, which shifts the pivot state $s$ in the polynomial expansion. For the pivot state $s \in \mathbb{R}$ and truncation order $P=5$, the PSC-linearized dynamical system takes the form
\begin{widetext}
\begin{align}
    \frac{d}{dt}
    \begin{bmatrix}
        y_0 \\ y_1 \\ y_2 \\ y_3 \\ y_4 \\ y_5
    \end{bmatrix} = 
    \begin{bmatrix}
        0 & 0 & 0 & 0 & 0 & 0 \\
        b &-k & 0 &-a & 0 & 0 \\
        0 & 2b&-2k& 0 &-2a& 0 \\
        0 & 0 & 3b&-3k& 0 &-3a\\
        4as^6 & -24s^5 & 60as^3 & -4k+80as^3&-4k+60as^2& -24as \\
        30as^2 & -175as^6 & 420as^5 & -525as^4 & 5b+350as^3&-5k-105as^2
    \end{bmatrix}
    \begin{bmatrix}
        y_0 \\ y_1 \\ y_2 \\ y_3 \\ y_4 \\ y_5
    \end{bmatrix}.
    \label{eq:exp1PSC}
\end{align}

\end{widetext}
We can observe that only the coefficients in the bottom two rows, which are associated with the time evolution of the two highest-order variables $\{y_4, y_5\}$, differ from the Carleman-linearized matrix in Eq.~(\ref{eq:exp1CL}). This is because the PSC method can be considered as adjusting the finite-dimensional truncation with respect to Carleman linearization. 

The PSC method requires the selection of an appropriate pivot state. Previous studies have suggested that the pivot state should be as close to the current state as possible. In our case, although the sign of $u(t)$ is always expected to be positive because the sign of the external force is positive, the value of $u(t)$ cannot be determined before the simulation. Thus, we set $s=0.01$ as a small positive value that can be selected without any prior analysis. Alternatively, if we can simulate the time-evolution equation using classical methods when nonlinear terms are ignored, the pivot state can be determined from these simulation results. There is scope for improvement in the determination of an appropriate pivot state based on the nature of the problem.

Fig.~2(e) presents the simulation results based on the proposed method and PSC linearization. Unlike Carleman linearization, the solution obtained using the PSC method maintains convergence even when $b\geq 1.2$ and agrees closely with the exact solution for all $b$ values. This implies that the performance of the proposed method can be improved by appropriately selecting an exact linearization method for the obtained nonlinear time evolutions.

\begin{figure*}[p]
\includegraphics[width=1.0\textwidth]{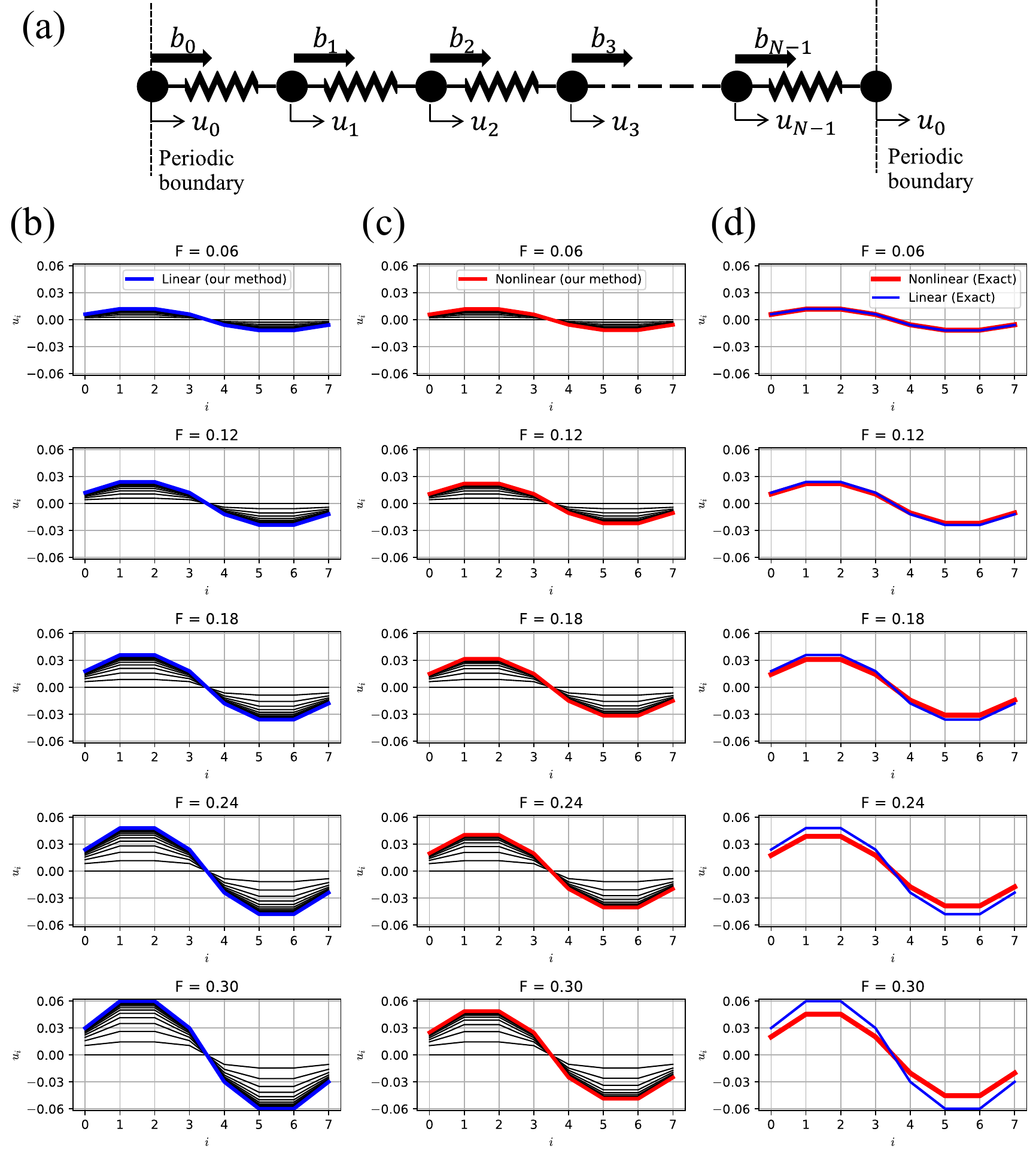}
\caption{Results for a one-dimensional nonlinear chain--spring system.
(a) Target chain--spring system with periodic boundary conditions. (b) Displacements obtained using our method with Carleman linearization for the linear spring case (blue). Intermediate trajectories are also plotted (black). (c) Displacements obtained using our method with Carleman linearization for the nonlinear spring case (red).
(d) Exact solution for the linear spring (blue) and nonlinear spring (red).
}
\label{fig:experimentY}
\vspace{0.5cm}
\end{figure*}

\subsection{One-dimensional chain of nonlinear springs}

Next, we consider a one-dimensional chain of multiple masses, as shown in Fig.~3(a). 
Let the number of point masses be $N$. The $i$-th mass is located at $x=i+u_i$, where $u_i$ denotes its displacement from the natural position. We impose periodic boundary conditions with length $N$ and connect all adjacent masses with identical nonlinear springs governed by Eq.~(\ref{eq:springforce}). All springs are at their natural length when $u_i=0$; therefore, the natural length of all springs is 1. 

Let the external forces $\bm{b}=(b_0,b_1,\cdots,b_{N-1})$ act on each mass. Our objective is to determine the equilibrium displacement $\bm{u}=(u_0,u_1,\cdots,u_{N-1})$. As in the previous section, the total potential energy of this system can be expressed as
\begin{align}
    U_\mathrm{int}(\bm{u}) = \sum^{N-1}_{i=0} \left[ \frac{1}{2}k(u_i-u_{i+1})^2 + \frac{1}{4}a(u_i-u_{i+1})^4 -b_i u_i \right],
\end{align}
where we use the notation $u_{N}=u_0$. Thus, the time-evolution equation that yields the equilibrium state is
\begin{align}
    \frac{du_i}{dt}
    =& -k\left[ (u_i-u_{i+1})+(u_i-u_{i-1}) \right] \nonumber \\
    &- a \left[ (u_i-u_{i+1})^3+(u_i-u_{i-1})^3 \right] +b_i
    \label{eq:exp2ev}
\end{align}
for all $i$. 

To simplify the subsequent analysis, we rewrite Eq.~(\ref{eq:exp2ev}) as
\begin{align}
    \frac{d\bm{u}}{dt} = F_0 \bm{u}^{\otimes 0} + F_1 \bm{u}^{\otimes 1} + F_3 \bm{u}^{\otimes 3}
    \label{eq:exp2ev_vec}
\end{align}
where 
\begin{align}
    F_0 &= \left[\bm{b}\right] \in \mathbb{R}^{N\times 1},\\
    F_1 &= k(S+S^T-2I) \in \mathbb{R}^{N\times N},\\
    F_3 &= aV\left\{(S^T-I)^{\otimes3}-(I-S)^{\otimes3}\right\} \in \mathbb{R}^{N\times N^3}.
\end{align}
Here, we use the cyclic shift matrix $S$ defined as $(S\bm{u})_i = u_{i+1}$ or equivalently, $(S^T\bm{u})_i = u_{i-1}$, and the special projector matrix $V$ defined as $(V\bm{u}^{\otimes 3})_i = u_i^3$. The equivalence of Eqs.~(\ref{eq:exp2ev}) and (\ref{eq:exp2ev_vec}) is presented in Appendix~B.

Next, we apply Carleman linearization. We introduce variables $\bm{y}_p = \bm{u}^{\otimes p}$. Taking the time derivative of the variables $\bm{y}_p$, we obtain the following infinite-dimensional linear time-evolution equation: 
\begin{align}
    \frac{d \bm{y}_p}{dt} = C(F_0,p) \bm{y}_{p-1} + C(F_1,p) \bm{y}_{p} + C(F_3,p) \bm{y}_{p+2}, 
    \label{eq:teveq_exp2}
\end{align}
where we define the matrix-valued function
\begin{align}
    C(F,p) = \sum_{v=0}^{p-1} \underbrace{I\otimes\cdots\otimes\ I}_{v \ \mathrm{times}}\otimes F \otimes \underbrace{I\otimes\cdots\otimes\ I}_{p-1-v \ \mathrm{times}}.
\end{align}
For example, truncating this time-evolution equation to the order $P=5$ yields
\begin{widetext}
\begin{align}
    \frac{d}{dt}
    \begin{bmatrix}
        y_0 \\ \bm{y}_1 \\ \bm{y}_2 \\ \bm{y}_3 \\ \bm{y}_4 \\ \bm{y}_5
    \end{bmatrix} = 
    \begin{bmatrix}
        \bm{O} & \bm{O} & \bm{O} & \bm{O} & \bm{O} & \bm{O} \\
        C(F_0,1)&C(F_1,1)& \bm{O} &C(F_3,1)& \bm{O} & \bm{O} \\
        \bm{O} &C(F_0,2)&C(F_1,2)& \bm{O} &C(F_3,2)& \bm{O} \\
        \bm{O} & \bm{O} &C(F_0,3)&C(F_1,3)& \bm{O} &C(F_3,3)\\
        \bm{O} & \bm{O} & \bm{O} &C(F_0,4)&C(F_1,4)& \bm{O} \\
        \bm{O} & \bm{O} & \bm{O} & \bm{O} &C(F_0,5)&C(F_1,5)
    \end{bmatrix}
    \begin{bmatrix}
        y_0 \\ \bm{y}_1 \\ \bm{y}_2 \\ \bm{y}_3 \\ \bm{y}_4 \\ \bm{y}_5
    \end{bmatrix},
    \label{eq:exp2CL}
\end{align}
\end{widetext}
where $\bm{O}$ is a zero matrix.

To evaluate the effectiveness of this linearization, we perform classical numerical simulations of Eq.~(\ref{eq:exp2CL}). We use the number of masses $N=8$ and the same spring parameters as in the previous section: $k=10, a=3000$. As external forces, we apply $+F$ to the left half of the chain and $-F$ to the right half of the chain, or
\begin{align}
    b_0 = b_1 = b_2 = b_3 =F, \ b_4 = b_5 = b_6 = b_7 =-F,
\end{align}
where we perform simulations for different values of $F$ from 0.00 to 0.30. We then obtain the numerically integrated results of Eq.~(\ref{eq:exp2CL}) on $t=1$.

The simulation results are shown as the red curves in Fig.~3(c). In this figure, the displacement of the mass $u_i$ indexed by the number $i$ on the horizontal axis is plotted on the vertical axis.
The black lines represent the transient behavior of the states up to $t=1$. For comparison, the red curves in Fig.~3(d) show the exact solution obtained by directly solving the nonlinear equilibrium equation $\nabla_u U(\bm{u})=0$ using the classical solver. In addition, we include the same simulation results, except for the use of linear springs (i.e., $k=10, a=0$). For the linear case, the blue curves in Fig.~3(b) shows the results obtained using the proposed method, and the blue curves in Fig.~3(d) shows the exact solution obtained using the classical solver.

The results obtained using the proposed method (indicated in red in Fig.~3(c)) agree with the exact nonlinear solution (red in Fig.~3(d)) for all $F$. A comparison between the nonlinear (red) and linear (blue) results of the exact simulation in Fig.~3(d) shows that the responses are nearly identical up to $F=0.12$, indicating that spring nonlinearity appears beyond this range. The results obtained using the proposed method (Fig.~3(c)) accurately reproduce these transition from linear to nonlinear, demonstrating that our method is also effective for this nonlinear chain system.

We can also estimate the query complexity to simulate the truncated linearized dynamics for this chain system in general, $N$ and $P$. As a generalization of Eq.~(\ref{eq:exp2CL}), we can express the Carleman-linearized time-evolution equation as
\begin{align}
    \frac{d\bm{y}}{dt} = K \bm{y}, 
    \label{eq:dydt_Ky}
\end{align}
where $\bm{y}=(\bm{y}_0,\cdots,\bm{y}_P)^T$, the $(i,j)$-block of $K$ is $C(F_{j-i+1},i)$ for $i>0$ and $j-i+1\in \{0,1,3\}$ and a zero matrix for the other case ($0 \leq i,j \leq P$). Earlier studies on HSs \cite{low2017optimal,low2019hamiltonian, low2025optimal} suggested that, given the block encoding of $K/\alpha$, the query complexity for simulating Eq.~(\ref{eq:dydt_Ky}) increases as 
\begin{align}
    \mathcal{O}(\alpha t\ln(1/\epsilon)).
    \label{eq:query_general_linear}
\end{align}
Here, $\alpha$ is the scaling factor for embedding the matrix $K$ into the block-encoding unitary matrix, $t$ is the length of time evolution, and $\epsilon$ is error. In best case, the scaling factor can be selected to $\alpha = \|K\|_2$. This makes the query complexity to simulate Eq.~(\ref{eq:dydt_Ky}) on quantum computer as
\begin{align}
    \mathcal{O}( P^{3/2}(\|\bm{b}\|+k+a)t \ln(1/\epsilon) ).
    \label{eq:qc_estimted}
\end{align}
Appendix~C provides proof of the above estimation.

This analysis reveals that the query complexity scales linearly with respect to the norm of the external force $\|\bm{b}\|$, linear and nonlinear coefficients of the spring $k,a$, and the truncation order of the power of three halves $P^{3/2}$. When the external force vector $\bm{b}$ is sparse, the norm is approximately constant. The coefficients $k$ and $a$, as well as the truncation order $P$, are also constant. Thus, the query complexity of this time evolution does not depend on the number of masses $N$ for any $t$. The theoretical bound of the required time-evolution length $t$, on which the overall complexity depends, is derived in Sec.~\ref{sec:theo}.

We can also estimate the required number of qubits. We assume that the additional ancilla qubits required for the block encoding of $K/\alpha$ and the HS are constant with respect to $N$. Because the dimension of $\bm{y}$ is
\begin{align}
    \mathrm{dim}(\bm{y}) = \sum_{p=1}^P N^p \leq 2 N^P = 2^{1+P \log_2 N},
\end{align}
the required number of qubits scales as $\mathcal{O}(P \log N)$. In other words, we require only a logarithmic number of quantum bits with respect to the size of the system $N$, indicating a quantum advantage over classical computers.

\begin{figure*}[t]
\includegraphics[width=1.0\textwidth]{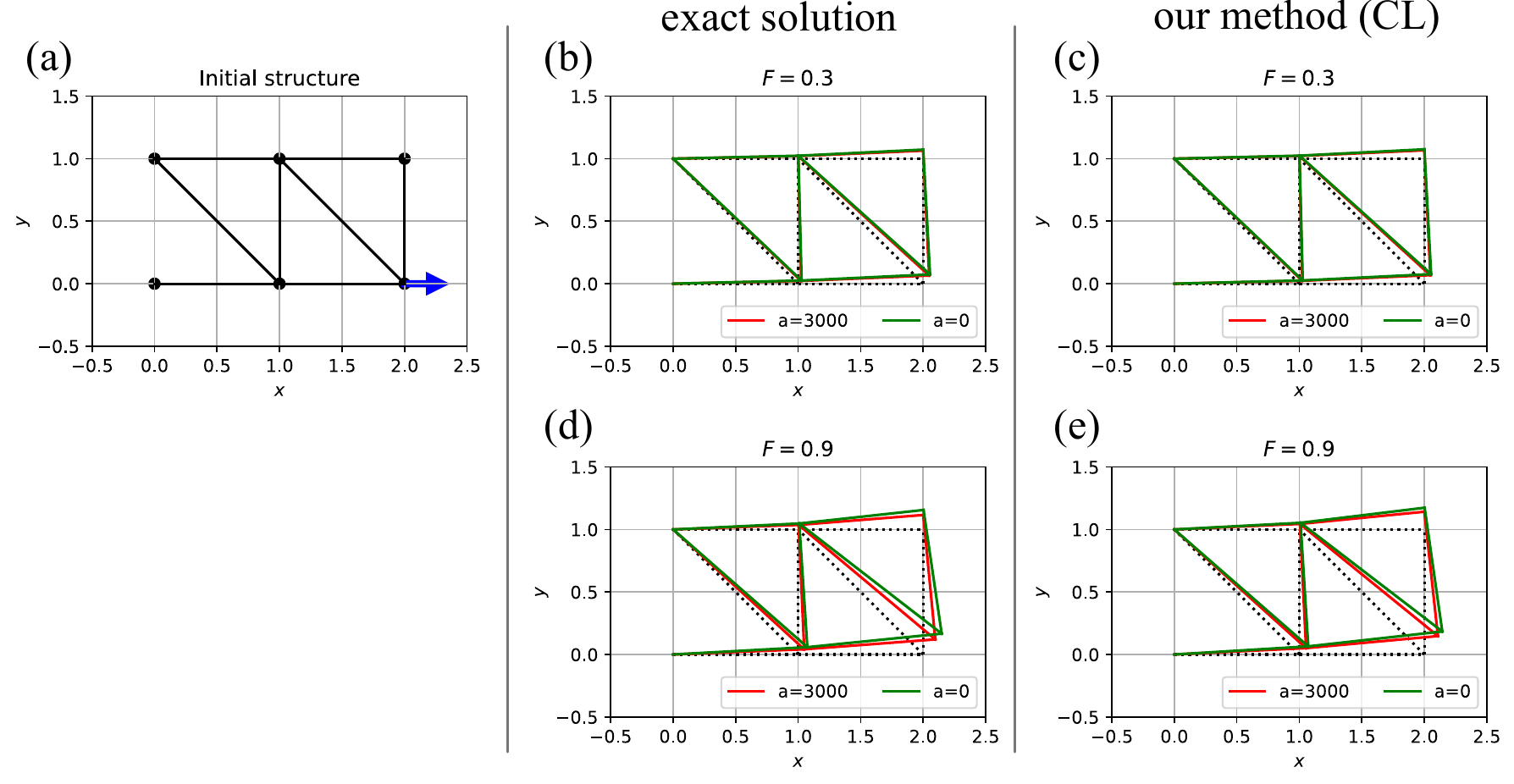}
\caption{Results for a two-dimensional truss system. (a) Target truss composed of masses (black points) and springs (black lines). The external force is indicated as a blue arrow with strength $F$. 
(b) Exact deformation at $F=0.3$ for linear springs (green) and nonlinear springs (red). (c) Deformation obtained using our method with Carleman linearization at $F=0.3$. (d) Exact deformation at $F=0.9$. (e) Deformation obtained using our method with Carleman linearization at $F=0.9$.}
\label{fig:experimentZ}
\end{figure*}

\subsection{Two-dimensional nonlinear truss structure}

As a final example, we consider a two-dimensional truss composed of nonlinear spring elements, as shown in Fig.~4(a). This example demonstrates that the proposed method can be applied to more practical systems.

Consider $N$ point masses in a two-dimensional space, connected by nonlinear springs between two masses $(i,j)\in G$ in set $G$. When all springs are at their natural lengths, we assume that the $i$-th mass is located at a fixed point $(x_i,y_i)$. The displacement relative to this natural position is denoted by $(u_i, v_i)$. Our objective is to obtain the displacement of all the points when each external force $(f_x^i, f_y^i)$ is applied to the $i$-th mass.  

We begin with the internal potential energy, which is the sum of the internal energies of all the springs in $G$. By using the same nonlinear spring as in Eq.~(\ref{eq:springforce}), the internal energy $U_{i,j}$ of the spring $(i,j)$ is
\begin{align}
    U_{ij} = \frac{k}{2}(\Delta L_{i,j})^2 + \frac{a}{4}(\Delta L_{i,j})^4,
    \label{eq:exp3_intene}
\end{align}
where we denote the length of the spring $\Delta L_{i,j}$. $\Delta L_{i,j}$ can be expressed as
\begin{align}
    \Delta L_{i,j} = \sqrt{ (x+u)^2 + (y+v)^2 } - L_0,
\end{align}
where we define $x=x_i-x_j$, $y=y_i-y_j$, $u=u_i-u_j$, $v=v_i-v_j$ and $L_0=\sqrt{x^2+y^2}$ for simplicity. 

We approximate $U_{i,j}$ as a polynomial function for modeling, similar to taking a first-order approximation for infinitesimal deformations. Note that this polynomial approximation is {\bf not} related to exact linearization methods such as Carleman linearization. Because the energy of this nonlinear spring includes a quartic term (related to $a$), we use a fourth-order polynomial approximation to maintain the material nonlinearity. Subsequently, the energy of the spring is approximated as $\hat{U}_{i,j} =$
\begin{align}
\frac{k}{2}\left(\frac{l_1^2}{L_0^2}+\frac{l_1 l_2}{L_0^2}-\frac{l_1^3}{L_0^4}+\frac{l_2^2}{4L_0^2}-\frac{3l_1^2l_2}{2L_0^4}+\frac{5l_1^4}{4L_0^6}\right) + \frac{a l_1^4}{4 L_0^4},
\label{eq:2denemodel}
\end{align}
where we set $l_1 = ux+vy$ and $l_2 = u^2+v^2$. The proof is provided in Appendix~D. Note that $l_1$ and $l_2$ have only linear and quadratic terms with respect to the variables, respectively.

Therefore, including the energy of the external forces, the total potential energy of system $U$ is
\begin{align}
U = \sum_{(i,j) \in G} \hat{U}_{ij} - \sum_{i=1}^N (f_x^i u_i+f_y^i v_i).
\label{eq:2t_allpot}
\end{align}
Subsequently, we can calculate the equilibrium displacement $\{(u_i,v_j)\}$ that satisfies $\nabla_{u_i}U=\nabla_{v_i}U=0$ by using our method.

We solve the equilibrium displacement of the truss with the initial structure shown in Fig.~4(a). All springs use $k=10$ and $a=3000$, which is consistent with the previous examples. To confirm the nonlinear effects, we also consider the purely linear case $k=10$ and $a=0$. In this initial structure, six masses (black dots) exist, but the masses at $(0,0)$ and $(0,1)$ are fixed. Therefore, these two masses are excluded from the state vector. An external force is applied only at the node located at $(2,0)$ with the force vector $(F,0)$ (blue arrow). We use $F=0.3$ and $0.9$ in this experiment. The nonlinear time-evolution equations obtained by our method are linearized using Carleman linearization and truncated at the order $P=5$. The solutions are obtained by numerically integrating the obtained truncated time evolution.

Fig.~4(c) and 4(e) show the results obtained for $F=0.3$ and $0.9$, respectively. For comparison, we calculate the exact solution by minimizing the total potential Eq.~(\ref{eq:2t_allpot}) using the classical nonlinear solver. Fig.~4(b) and 4(e) show the exact solutions for $F=0.3$ and $0.9$, respectively. In these four figures, the linear case ($k=10, a=0$) is plotted in green and the nonlinear case ($k=10, a=3000$) in red.

At $F=0.3$, the external force is sufficiently small, such that the response remained linear. This is confirmed by the coinciding of the exact solutions for the linear (green) and nonlinear (red) cases  in Fig.~4(b). The results obtained using the proposed method in Fig.~4(c) match these exact solutions correctly for both the linear and nonlinear cases. Thus, when the displacements remain in the linear region, the proposed method accurately predicts the response to an external force.

For $F=0.9$, the exact solution shown in Fig.~4(d) shows that the nonlinear case (red) exhibits reduced displacements relative to the linear case (green). Fig.~4(e) shows that the proposed method successfully reproduce these reduced displacements. However, the suppression of displacements is slightly milder than that in the exact solutions. This is due to the errors introduced by Carleman linearization, as confirmed by the first and second numerical experiments.

\section{Computational Complexity and Pre-initialization Technique}
\label{sec:theo}
In the previous section, we have applied the time-evolution conversion of the static equation and Carleman linearization to concrete nonlinear static problems and confirmed their numerical usefulness.
In this section, we present a theoretical evaluation of the computational complexity of the proposed method for solving nonlinear static problems.

For complexity evaluation, the most important aspect of our method is to estimate the time evolution required for the linear system obtained using Carleman linearization.
As indicated in Eq.~(\ref{eq:query_general_linear}), the query complexity of the quantum simulation of the linear system increases in proportion to the product of the spectral norm of the linearized matrix $K$ and time-evolution length $T$.
Among these quantities, the spectral norm of the linear matrix $K$ obtained through Carleman linearization can be bounded using the polynomial coefficients of the original nonlinear static equation and the truncation order of the Carleman linearization.
As shown in Sec.~\ref{sec:results} B, this bound can be evaluated concretely for each system.
However, the required time length $T$ depends on the initial condition $\bm{y}(0)$ and on the structure of the nonlinear dynamics; therefore, it must be evaluated separately.

In the following, we assume that the nonlinear static equation to be solved is 
\begin{align}
\bm{0}
=
\bm{b}+A\bm{x}+\eta\bm{g}(\bm{x}).
\label{eq:static_eq_pert}
\end{align}
Here, $\bm{g}(\bm{x})$ is a polynomial that contains neither constant nor linear terms and is defined as
\begin{align}
\bm{g}(\bm{x})
=
\sum_{d=2}^{D}F_d\bm{x}^{\otimes d}.
\label{eq:g_def}
\end{align}
We also assume $\bm{x},\bm{b}\in\mathbb{R}^N$, $A\in\mathbb{R}^{N\times N}$, $F_d\in\mathbb{R}^{N\times N^d}$, and $\eta\ge0$.

In addition, according to the requirements of the proposed method, we assume a potential function $U$ exists independently of $\eta\ge0$, such that the right-hand side of the above equation satisfies
\begin{align}
\bm{b}+A\bm{x}+\eta\bm{g}(\bm{x})
=
-\nabla_{\bm{x}}U(\bm{x}),
\label{eq:potential_exists}
\end{align}
and the assumptions of the proposed method are satisfied.
From this assumption, the linear part $A$ is a symmetric matrix, and the largest eigenvalue $\alpha_1$ of $A$ is necessarily negative.

Theoretical calculations based on these assumptions provide an estimate of the required time length $T$ that is valid when $\eta$ is not excessively large.
Leaving a detailed discussion, including the proof, to Appendix~E, the argument can be summarized as follows:
Based on the proposed method, we transform this static equation into a nonlinear time-evolution equation and apply Carleman linearization with truncation order $P$.
This yields the following finite-dimensional linear time-evolution equation:
\begin{align}
\frac{d\bm{y}}{dt}=K\bm{y}.
\label{eq:linear_evolution_K}
\end{align}
Here, $K$ is the linear matrix obtained through Carleman linearization.
It can be expressed as
\begin{align}
K=K^{(0)}+\eta K^{(f)},
\label{eq:K_decomposition}
\end{align}
where $K^{(0)}$ is determined only by the linear part of the original nonlinear static equation ($\bm{b}$ and $A$) and linearization order $P$, and $K^{(f)}$ is determined only by the nonlinear part ($F_d$, $D\ge d\ge2$) and linearization order $P$.
When $\bm{y}$ evolves according to Eq.~(\ref{eq:linear_evolution_K}), the proposed method implies that $\bm{y}$ reaches the final state $\bm{y}(\infty)$.
Therefore, we define the error vector between the state $\bm{y}(t)$ at time $t$ and the final state $\bm{y}(\infty)$ as
\begin{align}
\bm{z}(t)=\bm{y}(t)-\bm{y}(\infty),
\label{eq:z_def}
\end{align}
and define $T$ as the time at which the norm of the error vector falls below the target error threshold $\Delta>0$.
Equivalently, $T$ is a quantity that satisfies
\begin{align}
t\ge T
\quad\Rightarrow\quad
\|\bm{z}(t)\|\le\Delta.
\label{eq:T_definition}
\end{align}

Thus, $T$ can be bounded as
\begin{align}
T
\le
T_{\mathrm{bound}}
:=
\frac{
\log\|\bm{z}(0)\|_{\rho}
+
\log\left(\frac{1}{\Delta}\right)
}{
-\mu_{\rho}
}.
\label{eq:T_bound_main}
\end{align}
Here, $\rho\ge1$ is a parameter that adjusts weights to different degrees during Carleman embedding.
The quantity $\|\cdot\|_{\rho}$ is a suitably defined weighted norm, and $\mu_{\rho}$ is a constant that depends only on $\rho,A,\bm{b},\eta,\bm{g}(\bm{x})$, and $P$; their detailed definitions are given in Appendix~E.

Equation~(\ref{eq:T_bound_main}) estimates the required computation time using the proposed method.
This equation shows that the required time is proportional to the sum of the logarithms of the inverse target error threshold $\Delta$ and the logarithm of the weighted norm $\|\bm{z}(0)\|_{\rho}$ of the error at the initial state.
The proportionality factor is determined by the inverse of the convergence rate $\mu_{\rho}$, which includes the effects of both linear and nonlinear terms.
The quantity $\mu_{\rho}$ in Eq.~(\ref{eq:T_bound_main}) depends only on $\rho,A,\bm{b},\eta,\bm{g}(\bm{x})$, and $P$.
That is, as $\mu_{\rho}$ approaches zero, the convergence to the final state becomes slower and the required time length $T$ increases.
For a fixed $\rho$, a stronger nonlinear term increases $\mu_{\rho}$ toward zero, thereby reducing the effective decay rate $-\mu_{\rho}$.
This reflects the fact that nonlinearity can slow the convergence.

Because $\mu_{\rho}$ is determined solely by the original nonlinear static equation, it cannot be improved when the system is fixed.
However, $\|\bm{z}(0)\|_{\rho}$ depends on the initial state used when applying the proposed method; therefore, it has a certain degree of freedom.
A particularly important point is that, if an initial state $\bm{y}(0)$ close to the final state $\bm{y}(\infty)$ can be selected, then $\|\bm{z}(0)\|_{\rho}$ becomes small, and the term $\log(\|\bm{z}(0)\|_{\rho})$ can substantially reduce the required time length $T$.

Therefore, in practical applications of the proposed method, the key question is how to select, in a problem-dependent manner, an initial state $\bm{y}(0)$ that is close to the final state $\bm{y}(\infty)$.
We call this strategy \emph{pre-initialization}.
In particular, when the nonlinear term is small, namely, when $\eta$ is small, an effective choice is expected to be the Carleman-embedded vector of the solution $\bm{x}_*$ of
\begin{align}
A\bm{x}_*+\bm{b}=\bm{0},
\end{align}
namely
\begin{align}
\bm{y}_*
=
\left(
1,\bm{x}_*,
\bm{x}_*^{\otimes2},
\ldots,
\bm{x}_*^{\otimes P}
\right)^T.
\label{eq:y_star_preinit}
\end{align}
This vector $\bm{y}_*$ is expected to be close to the stationary state of the truncated Carleman-linearized dynamics.
Based on a perturbative theoretical analysis, the norm of the error vector at the initial time for an initial state $\bm{y}(0)$ can be calculated as
\begin{align}
\|\bm{z}(0)\|_{\rho}
=
\|\bm{y}_*-\bm{y}(0)\|_{\rho}
+O(\eta)
\label{eq:z0_preinit}
\end{align}
(See Appendix~E for the proof).
This implies that, if a state different from $\bm{y}_*$ is used as the initial state $\bm{y}(0)$, then $\|\bm{z}(0)\|_{\rho}$ has a constant dependence that is independent of $\eta$.
This provides a theoretical justification for selecting $\bm{y}(0)=\bm{y}_*$ when the structure of $\bm{g}(\bm{x})$ is unknown.

Next, we examine the behavior of the time estimate $T_{\mathrm{bound}}$ in the setting of a one-dimensional single nonlinear spring in Sec.~\ref{sec:results} A.
The target time-evolution equation is defined in Eq.~(\ref{eq:exp1CL}) of Sec.~\ref{sec:results} A.
Here, as in Sec.~\ref{sec:results} A, we set $k=10$, $b=0.2$, $P=5$, and $\Delta=10^{-8}$.
We vary the coefficient $a$, which represents the strength of spring nonlinearity, and compare the time $T_{\mathrm{hit}}$ at which the error threshold $\Delta$ is reached in the numerical experiments with the theoretical estimate $T_{\mathrm{bound}}$.
We also compare the case using $u(0)=0$ as the initial value with the pre-initialization case using the linear solution $u(0)=x_*=b/k=0.02$.

\begin{figure}[t]
\includegraphics[width=\columnwidth]{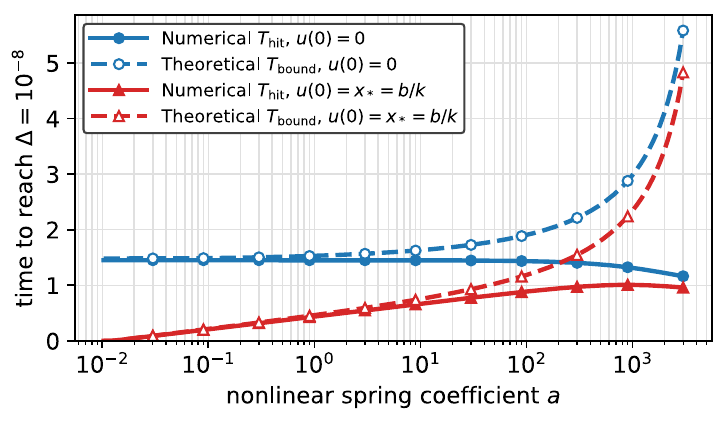}
\caption{Comparison of the numerical hitting time $T_{\mathrm{hit}}$ and theoretical upper bound $T_{\mathrm{bound}}$ for the single nonlinear spring problem.}
\label{fig:preinit_time_bound}
\end{figure}

Fig.~\ref{fig:preinit_time_bound} shows a comparison between $T_{\mathrm{hit}}$ and $T_{\mathrm{bound}}$ as functions of the coefficient $a$, which represents the strength of the spring nonlinearity.
The horizontal axis represents $a$, and the vertical axis represents the time required for the error to fall below the threshold $\Delta=10^{-8}$.
The solid lines represent $T_{\mathrm{hit}}$ computed in the numerical experiments, and the dashed lines represent the theoretical estimate $T_{\mathrm{bound}}$.
The blue lines represent the case $u(0)=0$, whereas the red lines represent the pre-initialization case using the linear solution $u(0)=x_*=b/k$.
The points on each line indicate the values of $a$ actually evaluated.

Fig.~\ref{fig:preinit_time_bound} shows that, even when $a$ is varied, the theoretical estimate $T_{\mathrm{bound}}$ is always larger than the numerically obtained $T_{\mathrm{hit}}$ and therefore functions as a theoretical upper bound.
However, in the regime in which the nonlinearity is large, the theoretical estimate becomes much larger than the numerical result.
This is because the derivation of the theoretical estimate uses inequality bounds that become loose in a strongly nonlinear regime.
In addition, compared with the case $u(0)=0$, both $T_{\mathrm{hit}}$ and $T_{\mathrm{bound}}$ become shorter when pre-initialization is used.
This indicates that appropriate pre-initialization reduces the required time.

Moreover, the estimate does not require an eigenvalue analysis of the full matrix $K$ and can be computed only from the degree-block structure and polynomial coefficients of the original equation.
Therefore, it is considered as a practical estimate of the time length used in the complexity evaluation of the proposed method.

\section{CONCLUSION}
In this paper, we introduce a quantum algorithmic framework for solving nonlinear equilibrium equations derived from variational principles. The central concept is to reformulate the equilibrium condition as nonlinear gradient-flow dynamics and transform the resulting time-evolution equation into a linear dynamical system through exact linearization techniques such as Carleman linearization and the pivot switching Carleman (PSC) method. This formulation enables the use of fault-tolerant quantum algorithms for linear dynamical simulations, including quantum linear system algorithms and HS methods, thereby providing a route toward the quantum acceleration of nonlinear equilibrium analysis.

The key advantage of this formulation is that it connects nonlinear equilibrium problems to a class of linear dynamical simulations whose resource requirements can be analyzed within the existing framework of quantum algorithms. For the nonlinear spring and truss systems studied in this paper, we demonstrate that truncated linearized dynamics can accurately reproduce the nonlinear equilibrium states. In particular, PSC linearization significantly improves the stability of truncated dynamics, enabling reliable simulations even in regimes where conventional Carleman linearization exhibits divergence.

We further analyze the computational resources required for the proposed approach. The query complexity for simulating the truncated linearized dynamics scales linearly with the evolution time and depends on the truncation order and physical parameters of the system, while remaining independent of the system size under the assumed access model. We also show that the number of qubits required to represent the Carleman-embedded state increases logarithmically with the number of degrees of freedom. These results indicate that quantum computers offer substantial memory advantages over classical approaches for large-scale nonlinear equilibrium problems. The complexity analysis presented here focuses on the simulation of linearized dynamics. The additional costs associated with state preparation and measurement depend on the specific implementation and remain an important topic for future investigation.

A central challenge in applying dynamical approaches to equilibrium problems is estimating the evolution time required to reach a stationary solution. To address this problem, we introduce a pre-initialization strategy based on the solution of the corresponding linearized equilibrium problem. Under a weak-nonlinearity assumption, we show that initializing the system near the linear solution reduces the initial error and substantially shortens the convergence time of the gradient-flow dynamics. This analysis provides a theoretical foundation for selecting effective initial states and highlights the importance of incorporating classical approximations when designing quantum algorithms for nonlinear problems.

From a broader perspective, the proposed framework can be interpreted as a quantum approach to nonlinear energy minimization. Because the equilibrium states derived from variational principles correspond to the stationary points of an energy functional, the gradient-flow formulation effectively converts nonlinear optimization problems into linearizable dynamical systems. This viewpoint suggests that the method may provide a new route toward the quantum acceleration of nonlinear optimization tasks beyond traditional linear-system-based quantum algorithms.

The applicability of the proposed approach extends beyond structural mechanics. Nonlinear equilibrium equations derived from variational principles appear in many areas of science and engineering, including fluid dynamics, electromagnetics, reaction–diffusion systems, and phase-field models. In these settings, the combination of gradient-flow reformulation, exact linearization, and quantum dynamical simulation may offer a promising pathway for analyzing complex nonlinear phenomena in future fault-tolerant quantum computers.

However, several important directions for future research remain unexplored. These include a more detailed analysis of the truncation errors in the linearization procedure, improved strategies for selecting pivot states in PSC linearization, and a comprehensive evaluation of the state preparation and readout costs in practical quantum implementations. Further investigation of hybrid classical–quantum strategies, such as the use of classical approximations for pre-initialization, may lead to more efficient algorithms for large-scale nonlinear systems.

We hope that the framework presented in this paper will stimulate further research on quantum algorithms for nonlinear problems and contribute to bridging the gap between nonlinear physical modeling and quantum computational methods.

\begin{acknowledgments}
We would like to thank Editage (www.editage.jp) for English language editing.
This paper is partially based on results obtained from a project, Council for Science, Technology and Innovation (CSTI), Cross-ministerial Strategic Innovation Promotion Program (SIP), ``Promoting the application of advanced quantum technology platforms to social issues'' (Funding agency : QST). This work was partially supported by JST [Moonshot R\&D] [Grant Number JPMJMS256J].
K.Z.T. was partially supported by JSPS KAKENHI Grant Number JP 25K01180.
\end{acknowledgments}

\bibliography{apssamp}

\appendix
\section{Simple proof for the arrival to the solution}
Consider the time-evolution equation 
\begin{align}
    \frac{d\bm{u}}{dt} = - \bm{f}(\bm{u}).
    \label{eq:fdynamicsnonlin_APPENDIX}
\end{align}
As shown in Eq.~\ref{eq:fdynamicsnonlin}, $\bm{f}$ satisfies 
\begin{align}
    \bm{f}(\bm{u}) = \bm{\nabla}_{\bm{u}} U
\end{align}
where (1) $\nabla_{\bm{u}}U$ is locally Lipschitz continuous anywhere and (2) $U(\bm{u})\rightarrow \infty$ when $\|\bm{u}\| \rightarrow \infty$.

According to the Picard–Lindelöf theorem \cite{coddington1955theory}, because $\bm{f}$ is locally Lipschitz continuous, the trajectory of $\bm{u}(t)$ is unique to any initial state $\bm{u}(0)$. The condition locally Lipschitz continuous is also implies that $U$ is a continuous function. The continuity of $U$ and the condition $U(\bm{u})\rightarrow \infty$ as $\|\bm{u}\| \rightarrow \infty$ imply that $U$ has a global minimum. 

Because
\begin{align}
    \frac{dU}{dt}(\bm{u}) = -\bm{\nabla}_{\bm{u}} U \cdot \frac{d\bm{u}}{dt}
    = - \| \bm{\nabla}_{\bm{u}} U\|^2 \leq 0,
\end{align}
$U(t)$ decreases monotonically. Combined with the fact $U$ has a global minimum value, the limit of $U(t\rightarrow \infty)$ exists, and its value is finite. This implies that 
\begin{align}
    \int_{t=0}^{\infty} \| \bm{\nabla}_{\bm{u}} U\|^2
    = U(t=0) - U(t\rightarrow \infty)
\end{align}
is finite, resulting
\begin{align}
    \lim_{t\rightarrow \infty} \| \bm{\nabla}_{\bm{u}} U\| = 0.
\end{align}

\section{Equivalence of Eq.~(\ref{eq:exp2ev}) and Eq.~(\ref{eq:exp2ev_vec})}

We expand the right-hand side of the following equation:
\begin{align}
    \frac{d\bm{u}}{dt} = F_0 \bm{u}^{\otimes 0} + F_1 \bm{u}^{\otimes 1} + F_3 \bm{u}^{\otimes 3}
    \label{eq:exp2ev_vec_APENDIX}
\end{align}
 (identical to Eq.~(\ref{eq:exp2ev_vec})) element by element and verified to match Eq.~(\ref{eq:exp2ev}).

For the first term, we obtain
\begin{align}
    (F_0 \bm{u}^{\otimes 0})_i = (\bm{b})_i = b_i.
\end{align}
For the second term, we obtain
\begin{align}
    & (F_1 \bm{u}^{\otimes 1})_i \nonumber \\
    =& k((S+S^T-2I)\bm{u})_i \nonumber \\
    =& k((S\bm{u})_i+(S^T\bm{u})_i-2(\bm{u})_i \nonumber \\
    =& k(u_{i+1} +u_{i-1} - 2u_i) \nonumber \\
    =& -k[(u_i-u_{i+1})+(u_i-u_{i-1})].
\end{align}
For the third term, we obtain
\begin{align}
    & (F_3 \bm{u}^{\otimes 3})_i \nonumber \\
    &=\left( aV\left\{(S^T-I)^{\otimes3}-(I-S)^{\otimes3}\right\} \bm{u}^{\otimes 3}\right)_i \nonumber \\
    &=  a\left( V
        (S^T\bm{u}-\bm{u})^{\otimes 3}
        \right)_i
        -
        a\left( V 
        (\bm{u}-S\bm{u})^{\otimes 3}
        \right)_i
    \nonumber \\
    & =
    a [(S^T\bm{u}-\bm{u})_i]^3 - a [(\bm{u}-S\bm{u})_i]^3
    \nonumber \\
    & = - a \left[ (u_i-u_{i+1})^3+(u_i-u_{i-1})^3 \right]
\end{align}

Therefore, the sum of the three terms is equal to Eq.~(\ref{eq:exp2ev}).

\section{Estimated query complexity of Eq.~(\ref{eq:dydt_Ky})}
We consider the case in which we have block encoding of $K/\alpha$, where $\alpha = \|K\|_2$. Herer, because the query complexity for simulating Eq.~(\ref{eq:dydt_Ky}) increases as $\mathcal{O}(\alpha t)$, we must estimate the size of $\alpha=\|K\|_2$ for a given $P, a, k, \bm{b}$.

First, we obtain the upper bound
\begin{align}
    \|K\|_2^2 \leq& \sum_{p=1}^{P} \|C(F_0,p)\|_2^2 \nonumber \\
    &+
    \sum_{p=1}^{P} \|C(F_1,p)\|_2^2 +
    \sum_{p=1}^{P-2} \|C(F_3,p)\|_2^2,
\end{align}
where we use the inequality $\|[X \ Y]\|^2_2 \leq \|X\|^2_2 + \|Y\|^2_2$ multiple times.

Next, we evaluate the upper bound of the spectral norm of $C(F,p)$ as
\begin{align}
    \|C(F,p)\| &= \left\|\sum_{v=0}^{p-1} \underbrace{I\otimes\cdots\otimes\ I}_{v \ \mathrm{times}}\otimes F \otimes \underbrace{I\otimes\cdots\otimes\ I}_{p-1-v \ \mathrm{times}}.\right\|_2 \nonumber\\
    &\leq \sum_{v=0}^{p-1} \left\| \underbrace{I\otimes\cdots\otimes\ I}_{v \ \mathrm{times}}\otimes F \otimes \underbrace{I\otimes\cdots\otimes\ I}_{p-1-v \ \mathrm{times}}.\right\|_2 \nonumber\\
    &= \sum_{v=0}^{p-1} \|F\|_2 = p\|F\|_2,
\end{align}
where we use the inequality $\|X+Y\|_2\leq\|X\|_2+\|Y\|_2$ and the equality $\|X\otimes Y\|_2=\|X\|_2\|Y\|_2$.
Thus, substituting Eq.~(C1) to Eq.~(C2), we obtain
\begin{align}
    \|K\|_2^2 \leq& \sum_{p=1}^{P} p^2\|F_0\|^2_2 +
    \sum_{p=1}^{P} p^2\|F_1\|^2_2 +
    \sum_{p=1}^{P-2} p^2\|F_3\|^2_2. \nonumber \\
    \leq& P^3 (\|F_0\|^2_2+\|F_1\|^2_2+\|F_3\|^2_2)
\end{align}

We also evaluate the upper bound of the spectral norm of $F_0, F_1, F_3$ as follows:
\begin{align}
    \|F_0\| = \|\bm{b}\|
\end{align}
\begin{align}
    \|F_1\| = \|k(S+S^T-2I)\| = k\|S+S^T-2I\|\leq 4k. 
\end{align}
\begin{align}
    \|F_3\| = \|aV\left\{(S^T-I)^{\otimes3}-(I-S)^{\otimes3}\right\}\| = \nonumber\\ a\|V\left\{(S^T-I)^{\otimes3}-(I-S)^{\otimes3}\right\}\|\leq 2\sqrt{10}a. 
\end{align}

Finally, by substituting Eqs.~(C4)--(C6) into Eq.~(C3), we obtain
\begin{align}
    \|K\|_2^2 \leq P^3 (\|\bm{b}\|^2+16k^2+40a^2). 
\end{align}
Therefore, we obtain the order of $\|K\|_2$ as
\begin{align}
\|K\|_2 = \mathcal{O}(P^{3/2}(\|\bm{b}\|+k+a)).
\end{align}
This immediately yields Eq.~(\ref{eq:qc_estimted}).

\section{Derivation of the energy model of two-dimensional nonlinear spring Eq.~(\ref{eq:2denemodel})}

The energy of the spring is defined as
\begin{align}
U_{ij} = \frac{k}{2} (\Delta L_{i,j})^2 + \frac{a}{4} (\Delta L_{i,j})^4,
\label{eq:2t_banene_APPENDIX}
\end{align}
where 
\begin{align}
\Delta L_{i,j} = \sqrt{(x+u)^2+(y +v)^2} - L_0
\end{align}
represents the extension of the spring. We can rewrite this spring extension as
\begin{align}
\Delta L_{i,j} &= L_0 \left( \sqrt{1 + \frac{2(ux+vy)+(x^2+y^2)}{L_0^2}} - 1 \right)\\
&= L_0 \left( \sqrt{1 + l} - 1 \right),
\end{align}
where we define $l=(2l_1+l_2)/L_0^2$, $l_1 = ux+vy$ and $l_2 = x^2+y^2$.

As the Taylor expansion of the function $\sqrt{1+p}$ is
\begin{align}
\sqrt{1+p}=1+\frac{1}{2}p - \frac{1}{8}p^2 + \frac{1}{16}p^3 - \frac{5}{128}p^4 + \cdots,       
\end{align}
the spring extension is approximated as
\begin{align}
\Delta L_{i,j} = L_0 \left( \frac{1}{2}l - \frac{1}{8}l^2 + \frac{1}{16}l^3 - \frac{5}{128}l^4 + \cdots \right).
\end{align}
By substituting this relationship into Eq.~(\ref{eq:2t_banene_APPENDIX}), we obtain
\begin{align}
U_{ij} &= \frac{kL_0^2 }{2} \left(\frac{1}{2}l-\frac{1}{8}l^2+\frac{1}{16}l^3+\cdots \right)^2 \\
&+ \frac{aL_0^4}{4} \left(\frac{1}{2}l-\frac{1}{8}l^2+\frac{1}{16}l^3 +\cdots \right)^4 \\
&= \frac{k}{2}\left(\frac{l_1^2}{L_0^2}+\frac{l_1 l_2}{L_0^2}-\frac{l_1^3}{L_0^4}+\frac{l_2^2}{4L_0^2}-\frac{3l_1^2l_2}{2L_0^4}+\frac{5l_1^4}{4L_0^6}\right) \\
&+ \frac{a l_1^4}{4 l_0^4} + \mathcal{O}((u+v)^5).
\end{align}
Ignoring terms of orders higher than 4, we obtain the following energy model: $\hat{U}_{ij} =$
\begin{align}
\frac{k}{2}\left(\frac{l_1^2}{L_0^2}+\frac{l_1 l_2}{L_0^2}-\frac{l_1^3}{L_0^4}+\frac{l_2^2}{4L_0^2}-\frac{3l_1^2l_2}{2L_0^4}+\frac{5l_1^4}{4L_0^6}\right) + \frac{a l_1^4}{4 L_0^4}.
\end{align}

\clearpage
\onecolumngrid
\section{Theoretical Evaluation of the Required Time}
\label{sec:apdx_time_theoretical}
In this appendix, we derive the time estimate $T_{\mathrm{bound}}$ used in the main text.

We assume that the nonlinear static equation to be solved is given by
\begin{align}
\bm{0}
=
\bm{b}+A\bm{x}+\eta\bm{g}(\bm{x}).
\label{eq:appE_static}
\end{align}
Here, $\bm{x},\bm{b}\in\mathbb{R}^N$ and $A\in\mathbb{R}^{N\times N}$.
The function $\bm{g}(\bm{x})$ is a real polynomial of degree $D$ that contains no terms of degree lower than two and is defined as
\begin{align}
\bm{g}(\bm{x})
=
\sum_{d=2}^{D}F_d\bm{x}^{\otimes d}.
\label{eq:appE_g_def}
\end{align}
Moreover, $\eta\ge0$ represents the strength of the nonlinear term.
Because the proposed method is also valid within the limit $\eta=0$, the linear part $A$ is symmetric.
We denote the largest eigenvalue of $A$ as $\alpha_1$, which satisfies
\begin{align}
\alpha_1
=
\lambda_{\max}(A)
<0.
\label{eq:appE_alpha1}
\end{align}

With the proposed method, this static equation is transformed into a nonlinear time-evolution equation:
\begin{align}
\frac{d\bm{x}}{dt}
=
\bm{b}+A\bm{x}
+\eta\sum_{d=2}^{D}F_d\bm{x}^{\otimes d}.
\label{eq:appE_nonlinear_dynamics}
\end{align}
We then apply Carleman linearization with truncation order $P$.
Let the Carleman variables be $\bm{y}_k=\bm{x}^{\otimes k}$, and the error vector corresponding to each degree is defined by
\begin{align}
\bm{z}_k(t)
=
\bm{y}_k(t)-\bm{y}_k(\infty),
\qquad
k=1,\ldots,P.
\label{eq:appE_zk_def}
\end{align}
By definition, $\bm{z}=(\bm{z}_0,\bm{z}_1,\ldots,\bm{z}_P)$.
Thus, for blocks of degree $k=1,\ldots,P$, the equation followed by the error vector can be expressed formally as
\begin{align}
\dot{\bm{z}}_k
=
A^{\oplus k}\bm{z}_k
+B^{\oplus k}\bm{z}_{k-1}
+\eta\sum_{d=2}^{D}
F_d^{\oplus k}\bm{z}_{k+d-1}.
\label{eq:appE_z_dynamics_block}
\end{align}
Here, components above the truncation order are set as $\bm{z}_k=\bm{0}$ for $k>P$.
We have also defined
\begin{align}
A^{\oplus k}
&=
\sum_{j=0}^{k-1}
I_N^{\otimes j}\otimes A\otimes I_N^{\otimes k-1-j},
\label{eq:appE_Aoplus}
\\
B^{\oplus k}
&=
\sum_{j=0}^{k-1}
I_N^{\otimes j}\otimes \bm{b}\otimes I_N^{\otimes k-1-j},
\label{eq:appE_Boplus}
\\
F_d^{\oplus k}
&=
\sum_{j=0}^{k-1}
I_N^{\otimes j}\otimes F_d\otimes I_N^{\otimes k-1-j}.
\label{eq:appE_Foplus}
\end{align}

Next, the weighted error vector is defined as
\begin{align}
\bm{\xi}_k=w_k\bm{z}_k.
\label{eq:appE_xi_def}
\end{align}
Here, the weight of the degree-$k$ error vector is selected as
\begin{align}
w_k=\rho^{k-1},
\qquad
\rho\ge1.
\label{eq:appE_weight_def}
\end{align}
The parameter $\rho$ is a numerical value that can be selected arbitrarily.
As $\bm{z}_k=w_k^{-1}\bm{\xi}_k$, we obtain
\begin{align}
\dot{\bm{\xi}}_k
=
A^{\oplus k}\bm{\xi}_k
+\frac{w_k}{w_{k-1}}B^{\oplus k}\bm{\xi}_{k-1}
+\eta\sum_{d=2}^{D}
\frac{w_k}{w_{k+d-1}}F_d^{\oplus k}
\bm{\xi}_{k+d-1}.
\label{eq:appE_xi_dynamics_pre}
\end{align}
Here,
\begin{align}
\frac{w_k}{w_{k-1}}=\rho,
\qquad
\frac{w_k}{w_{k+d-1}}=\rho^{-(d-1)}.
\label{eq:appE_weight_ratios}
\end{align}
Thus, the equation can be rewritten as
\begin{align}
\dot{\bm{\xi}}_k
=
A^{\oplus k}\bm{\xi}_k
+\rho B^{\oplus k}\bm{\xi}_{k-1}
+\eta\sum_{d=2}^{D}
\rho^{-(d-1)}F_d^{\oplus k}
\bm{\xi}_{k+d-1}.
\label{eq:appE_xi_dynamics}
\end{align}

We now define the weighted error energy and weighted norm as
\begin{align}
V(t)
:=
\|\bm{z}(t)\|_{\rho}^2
:=
\sum_{k=1}^{P}w_k^2\|\bm{z}_k(t)\|^2
=
\sum_{k=1}^{P}\|\bm{\xi}_k(t)\|^2.
\label{eq:appE_weighted_energy}
\end{align}
In the following, $\langle\cdot,\cdot\rangle$ denotes the Euclidean inner product.
The time derivative is
\begin{align}
\frac{dV}{dt}
=
2
\sum_{k=1}^{P}
\left\langle
\bm{\xi}_k,\dot{\bm{\xi}}_k
\right\rangle .
\label{eq:appE_dVdt_def}
\end{align}
Substituting the equation for the weighted error vector yields
\begin{align}
\frac{dV}{dt}
=
2\sum_{k=1}^{P}
\Bigg[
&
\left\langle
\bm{\xi}_k,
A^{\oplus k}\bm{\xi}_k
\right\rangle
\nonumber\\
&+
\left\langle
\bm{\xi}_k,
\rho B^{\oplus k}\bm{\xi}_{k-1}
\right\rangle
\nonumber\\
&+
\eta\sum_{d=2}^{D}
\left\langle
\bm{\xi}_k,
\rho^{-(d-1)}F_d^{\oplus k}
\bm{\xi}_{k+d-1}
\right\rangle
\Bigg].
\label{eq:appE_dVdt_expanded}
\end{align}
We now bound the terms on the right-hand side of Eq.~(\ref{eq:appE_dVdt_expanded}) separately.
For a linear term, the Rayleigh quotient estimate yields
\begin{align}
\left\langle
\bm{\xi}_k,
A^{\oplus k}\bm{\xi}_k
\right\rangle
\le
\lambda_{\max}\left(A^{\oplus k}\right)\|\bm{\xi}_k\|^2.
\label{eq:appE_rayleigh}
\end{align}
Moreover,
\begin{align}
\lambda_{\max}
\left(
A^{\oplus k}
\right)
=
k\alpha_1.
\label{eq:appE_lambda_Aoplus}
\end{align}
Therefore,
\begin{align}
\left\langle
\bm{\xi}_k,A^{\oplus k}\bm{\xi}_k
\right\rangle
\le
k\alpha_1\|\bm{\xi}_k\|^2
\label{eq:appE_linear_bound}
\end{align}
holds.

Next, we evaluate the term that contains $\bm{\xi}_{k-1}$ and originates from $\bm{b}$, and the term that contains $\bm{\xi}_{k+d-1}$ and originates from the nonlinear term.
First, the term originating from $\bm{b}$ can be bounded by
\begin{align}
\begin{split}
\left|
\left\langle
\bm{\xi}_k,
\rho B^{\oplus k}\bm{\xi}_{k-1}
\right\rangle
\right|
\le&
\|\bm{\xi}_k\|
\left\|
\rho B^{\oplus k}\bm{\xi}_{k-1}
\right\|
\\
\le&
\rho\|B^{\oplus k}\|
\|\bm{\xi}_k\|\|\bm{\xi}_{k-1}\|
\\
\le&
k\|\bm{b}\|\rho
\|\bm{\xi}_k\|\|\bm{\xi}_{k-1}\|.
\end{split}
\label{eq:appE_b_term_bound}
\end{align}
Here, we use the submultiplicativity of the norm and relation $\|B^{\oplus k}\|\le k\|\bm{b}\|$.
Using $2ab\le a^2+b^2$, we obtain
\begin{align}
k\|\bm{b}\|\rho
\|\bm{\xi}_k\|\|\bm{\xi}_{k-1}\|
\le
\frac{k\|\bm{b}\|\rho}{2}\|\bm{\xi}_k\|^2
+
\frac{k\|\bm{b}\|\rho}{2}\|\bm{\xi}_{k-1}\|^2.
\label{eq:appE_b_term_young}
\end{align}
Similarly, for the nonlinear term,
\begin{align}
&
\left|
\left\langle
\bm{\xi}_k,
\eta\rho^{-(d-1)}
F_d^{\oplus k}\bm{\xi}_{k+d-1}
\right\rangle
\right|
\nonumber\\
&\quad
\le
\eta k\|F_d\|\rho^{-(d-1)}
\|\bm{\xi}_k\|\|\bm{\xi}_{k+d-1}\|.
\label{eq:appE_f_term_bound}
\end{align}
Here, we use the relation $\|F_d^{\oplus k}\|\le k\|F_d\|$.
Similarly,
\begin{align}
&
\eta k\|F_d\|\rho^{-(d-1)}
\|\bm{\xi}_k\|\|\bm{\xi}_{k+d-1}\|
\nonumber\\
&\le
\frac{\eta k\|F_d\|\rho^{-(d-1)}}{2}
\|\bm{\xi}_k\|^2
+
\frac{\eta k\|F_d\|\rho^{-(d-1)}}{2}
\|\bm{\xi}_{k+d-1}\|^2
\label{eq:appE_f_term_young}
\end{align}
is obtained.

By combining the above estimates, we obtain
\begin{align}
\frac{dV}{dt}
\le&
2\sum_{k=1}^{P}
k\alpha_1\|\bm{\xi}_k\|^2
\nonumber\\
&+
2\sum_{k=2}^{P}
\left(
\frac{k\|\bm{b}\|\rho}{2}\|\bm{\xi}_k\|^2
+
\frac{k\|\bm{b}\|\rho}{2}\|\bm{\xi}_{k-1}\|^2
\right)
\nonumber\\
&+
2\sum_{d=2}^{D}
\sum_{k=1}^{P-d+1}
\left(
\frac{\eta k\|F_d\|\rho^{-(d-1)}}{2}
\|\bm{\xi}_k\|^2
+
\frac{\eta k\|F_d\|\rho^{-(d-1)}}{2}
\|\bm{\xi}_{k+d-1}\|^2
\right).
\label{eq:appE_dVdt_bound}
\end{align}
To obtain this expression as the coefficient of each degree-$j$ term $\|\bm{\xi}_j\|^2$, we define
\begin{align}
m_j(\rho)
=&\;
j\alpha_1
\nonumber\\
&+
\frac{1}{2}
\sum_{\substack{k=2,\ldots,P\\j=k}}
k\|\bm{b}\|\rho
+
\frac{1}{2}
\sum_{\substack{k=2,\ldots,P\\j=k-1}}
k\|\bm{b}\|\rho
\nonumber\\
&+
\frac{1}{2}
\sum_{d=2}^{D}
\sum_{\substack{k=1,\ldots,P-d+1\\j=k}}
\eta k\|F_d\|\rho^{-(d-1)}
\nonumber\\
&+
\frac{1}{2}
\sum_{d=2}^{D}
\sum_{\substack{k=1,\ldots,P-d+1\\j=k+d-1}}
\eta k\|F_d\|\rho^{-(d-1)}.
\label{eq:appE_mj_def}
\end{align}
Subsequently,
\begin{align}
\frac{dV}{dt}
\le
2
\sum_{j=1}^{P}
m_j(\rho)\|\bm{\xi}_j\|^2
\label{eq:appE_dVdt_mj}
\end{align}
is obtained.
We define
\begin{align}
\mu_{\rho}
=
\max_{1\le j\le P}
m_j(\rho).
\label{eq:appE_mu_def}
\end{align}
Thus, $m_j(\rho)\le\mu_{\rho}$ for all $j$, and $\|\bm{\xi}_j\|^2\ge0$.
Therefore,
\begin{align}
\sum_{j=1}^{P}
m_j(\rho)\|\bm{\xi}_j\|^2
\le
\mu_{\rho}
\sum_{j=1}^{P}
\|\bm{\xi}_j\|^2
=
\mu_{\rho}V(t)
\label{eq:appE_mu_bound}
\end{align}
holds.
Thus,
\begin{align}
\frac{dV}{dt}
\le
2\mu_{\rho} V(t)
\label{eq:appE_gronwall_ineq}
\end{align}
is obtained.
This differential inequality results in
\begin{align}
V(t)\le e^{2\mu_{\rho} t}V(0).
\label{eq:appE_V_bound}
\end{align}
It follows that
\begin{align}
\sqrt{V(t)}
\le
e^{\mu_{\rho} t}\sqrt{V(0)}.
\label{eq:appE_sqrtV_bound}
\end{align}

Moreover,
\begin{align}
\|\bm{z}(t)\|^2
\le
\sum_{k=1}^{P}w_k^2\|\bm{z}_k(t)\|^2
=
\sum_{k=1}^{P}\|\bm{\xi}_k(t)\|^2
=
V(t)
\label{eq:appE_unweighted_weighted}
\end{align}
holds.
In the first inequality, we use the fact that $w_k\ge1$ follows from $\rho\ge1$.
Therefore,
\begin{align}
\|\bm{z}(t)\|
\le
\sqrt{V(t)}
\label{eq:appE_z_sqrtV}
\end{align}
and
\begin{align}
\|\bm{z}(t)\|
\le
e^{\mu_{\rho} t}
\sqrt{V(0)}
\label{eq:appE_z_bound}
\end{align}
hold.
A sufficient condition for this to be no larger than $\Delta$ is
\begin{align}
e^{\mu_{\rho} t}
\sqrt{V(0)}
\le
\Delta.
\label{eq:appE_delta_condition}
\end{align}
When $\mu_{\rho}<0$, solving this condition for $t$ yields
\begin{align}
t
\ge
\frac{
\log\|\bm{z}(0)\|_{\rho}
+\log\left(\frac{1}{\Delta}\right)
}{
-\mu_{\rho}
}.
\label{eq:appE_t_condition}
\end{align}
Here, we use the definition of the weighted norm, which yields
\begin{align*}
V(0)=\|\bm{z}(0)\|_{\rho}^2.
\end{align*}
Thus, the upper bound of the time at which the norm of the error vector does not exceed the threshold $\Delta$ is obtained as
\begin{align}
T_{\mathrm{bound}}
:=
\frac{
\log\|\bm{z}(0)\|_{\rho}
+\log\left(\frac{1}{\Delta}\right)
}{
-\mu_{\rho}
}.
\label{eq:appE_T_bound}
\end{align}
Finally, because any choice of $\rho$ provides an upper bound on the required time length, we may select $\rho$ such that $T_{\mathrm{bound}}$ is minimized.
In the numerical experiment described in the main text, the plotted values were obtained by numerically optimizing $\rho$ in this manner.

Next, we present the initial-error estimate related to the pre-initialization used in the main text.
First, let $\bm{x}_*$ be the solution to the linear equation
\begin{align}
A\bm{x}_*+\bm{b}=\bm{0}.
\label{eq:appE_linear_solution}
\end{align}
We define its Carleman embedding as
\begin{align}
\bm{y}_*
=
\left(
1,\bm{x}_*,
\bm{x}_*^{\otimes2},
\ldots,
\bm{x}_*^{\otimes P}
\right)^T.
\label{eq:appE_y_star_def}
\end{align}
Restricting the Carleman vector to the nonconstant blocks $k=1,\ldots,P$, while absorbing the zeroth-order contribution into $\bm{c}$, the stationary equation after truncation can be expressed as
\begin{align}
\bm{0}
=
\bm{c}
+
\left(
K^{(0)}
+
\eta K^{(f)}
\right)
\bm{y}(\infty).
\label{eq:appE_truncated_stationary}
\end{align}
Here, $\bm{c}$ is a vector derived from the constant term $\bm{b}$.
When $\eta=0$, the nonlinear term does not exist. Therefore,
\begin{align}
\bm{0}
=
\bm{c}
+
K^{(0)}
\bm{y}_*
\label{eq:appE_linear_stationary}
\end{align}
holds.
Thus, for small $\eta$, we expand
\begin{align}
\bm{y}(\infty)
=
\bm{y}_*
+
\eta\bm{y}^{(1)}
+
O(\eta^2).
\label{eq:appE_y_infty_expansion}
\end{align}
Substituting this into the stationary equation yields
\begin{align}
&
\bm{0}
=
\bm{c}
+
\left(
K^{(0)}
+
\eta K^{(f)}
\right)
\left(
\bm{y}_*
+
\eta\bm{y}^{(1)}
+
O(\eta^2)
\right).
\label{eq:appE_stationary_substitution}
\end{align}
The zeroth-order term in $\eta$ is $\bm{c}+K^{(0)}\bm{y}_*$, which vanishes by Eq.~(\ref{eq:appE_linear_stationary}).
Extracting the first-order term yields
\begin{align}
K^{(0)}\bm{y}^{(1)}
+
K^{(f)}\bm{y}_*
=
\bm{0}.
\label{eq:appE_y_first_order}
\end{align}
Matrix $K^{(0)}$ is the truncated Carleman matrix for the linear problem, and it is invertible because it has only negative eigenvalues consisting of the sums of the eigenvalues of $A$.
Therefore,
\begin{align}
\bm{y}^{(1)}
=
-\left(K^{(0)}\right)^{-1}K^{(f)}\bm{y}_*
\label{eq:appE_y1_solution}
\end{align}
and
\begin{align}
\bm{y}(\infty)
=
\bm{y}_*
+
O(\eta)
\label{eq:appE_y_infty_close}
\end{align}
are obtained.
For a fixed $\rho$, this implies that
\begin{align}
\|\bm{y}(\infty)-\bm{y}_*\|_{\rho}
=
O(\eta)
\label{eq:appE_embedding_close}
\end{align}
holds.
However, because $\bm{z}(0)=\bm{y}(0)-\bm{y}(\infty)$, the triangle inequality yields
\begin{align}
\left|
\|\bm{z}(0)\|_{\rho}
-
\|\bm{y}_*-\bm{y}(0)\|_{\rho}
\right|
\le
\|\bm{y}(\infty)-\bm{y}_*\|_{\rho}
=
O(\eta).
\label{eq:appE_triangle}
\end{align}
Thus,
\begin{align}
\|\bm{z}(0)\|_{\rho}
=
\|\bm{y}_*-\bm{y}(0)\|_{\rho}
+O(\eta)
\label{eq:appE_z0_preinit}
\end{align}
holds.

\end{document}